\documentclass[%
 reprint,
superscriptaddress,
 amsmath,amssymb,
 aps,pre,hyphens,floatfix
]{revtex4-2}

\usepackage[]{amsmath}
\usepackage{mathrsfs}
\usepackage{mathtools} 
\usepackage{enumerate}
\usepackage[usenames,dvipsnames,svgnames,table]{xcolor}
\usepackage{IEEEtrantools}
\usepackage{graphicx}
\usepackage{array}
\usepackage{multirow}
\usepackage{verbatim}
\usepackage{soul}
\usepackage[normalem]{ulem}
\usepackage{tabularx}
\usepackage{booktabs,siunitx}
\usepackage[T1]{fontenc}
\usepackage{hyperref} 
\usepackage[hyphenbreaks]{breakurl}

\newcommand{\red}[1]{\textcolor{red}{#1}}

\newcommand{\mycomment}[1]{}

\begin{document}

\preprint{APS/123-QED}

\title{Revisiting and modeling power-law distributions in empirical outage data of power systems}

\author{B\'{a}lint Hartmann}
\email[]{hartmann.balint@ek-cer.hu}
\affiliation{Institute of Energy Security and Environmental Safety, Center for Energy Research, P.O. Box 49, H-1525 Budapest, Hungary}
\author{Shengfeng Deng}
\email[]{shengfeng.deng@ek-cer.hu}
\affiliation{Institute of Technical Physics and Materials Science, Center for Energy Research, P.O. Box 49, H-1525 Budapest, Hungary}

\author{G\'{e}za \'{O}dor}
\affiliation{Institute of Technical Physics and Materials Science, Center for Energy Research, P.O. Box 49, H-1525 Budapest, Hungary}

\author{Jeffrey Kelling}
\affiliation{Faculty of Natural Sciences, Chemnitz University of Technology, \\ Stra{\ss}e der Nationen 62,  09111 Chemnitz, Germany}
\affiliation{Department of Information Services and Computing,
Helmholtz-Zentrum Dresden-Rossendorf, P.O.Box 51 01 19, 01314 Dresden, Germany
}

\date{\today}

\begin{abstract}
The size distribution of planned and forced outages and following restoration times in power systems have been studied for almost two decades and has drawn great interest as they display heavy tails. Understanding of this phenomenon has been done by various threshold models, which are self-tuned at their critical points, but as many papers pointed out, explanations are intuitive, and more empirical data is needed to support hypotheses. In this paper, the authors analyze outage data collected from various public sources to calculate the outage energy and outage duration exponents of possible power-law fits. Temporal thresholds are applied to identify crossovers from initial short-time behavior to power-law tails. We revisit and add to the possible explanations of the uniformness of these exponents. By performing power spectral analyses on the outage event time series and the outage duration time series, it is found that, on the one hand, while being overwhelmed by white noise, outage events show traits of self-organized criticality (SOC), which may be modeled by a crossover from random percolation to directed percolation branching process with dissipation, coupled to a conserved density. On the other hand, in responses to outages, the heavy tails in outage duration distributions could be a consequence of the highly optimized tolerance (HOT) mechanism, based on the optimized allocation of maintenance resources.
\end{abstract}

\maketitle

\section{Introduction\label{sec:1}}

As the power sector undergoes an unprecedented transition, understanding the vulnerability of these systems receives more attention from the research community. Replacing conventional, dominantly fossil-fueled power plants with ones relying on variable sources such as solar and wind poses a number of challenges, mainly due to the appearing correlated spatio-temporal fluctuations, which in case of adverse conditions can lead to disturbances of various sizes. The size distribution of such events has been the focus of research for almost two decades, largely because it displays fat tails; a suitable candidate for fitting power laws. It is known that the severity of catastrophic events exhibits such behavior, even after removing extreme outliers \cite{hardle,chernobai}. The probability distribution of the number of people killed in natural disasters and man-made situations also shows power-law decay, with very similar exponent values. As presented in \cite{chatterjee}, the size distribution of deaths in man-made events (wars, battles, conflicts, terrorist attacks) decay with power-law tails, with exponents around 1.6. Similarly, the magnitudes of natural disasters (earthquakes, storms, wildfires, etc.) show exponents between 1.8 and 1.5. If such extreme events do follow a power law, they are not completely unexpected anymore; though they remain unpredictable. (These events were coined by Taleb as `gray swans' \cite{Taleb2007-qr}, in contrast to ‘black swans’, which can still occur due to the possible dynamics of the upper cutoff \cite{PhysRevResearch.4.033079}.)

The research community is actively working on the improvement of our understanding on how power outages can be forecasted  \cite{ABEDI2019153} and more specifically whether the risk of failure in power systems represents a specific case of the risk of system-wide breakdown in threshold activated disordered systems. Self-organized criticality (SOC), explained first by the Bak-Tang-Wieselfeld model \cite{bak1987}, is widely used for the modeling of such phenomena \cite{carreras2004}. In this regard, SOC is expected as the consequence of self-tuning to a critical point, which is determined by the competition of power consumption (behavioral properties) and available transmission capabilities (infrastructural properties) of the examined power system. In the early 2000's, another model has gained some attention in relation to the topic, namely highly optimized tolerance (HOT). HOT was coined by Carlson and Doyle, claiming that the model takes into consideration the fact that designs are developed and biological systems evolve in a manner that rewards successful strategies subject to a specific form of external stimulus \cite{PhysRevE.60.1412}. Following this original paper, Stubna and Fowler examined the HOT model using the data from the Western United States \cite{stubna2003application} and concluded that the model agrees closely if the outage event size is considered through the power loss but not for the number of consumers affected, thus the model of optimal resource distribution is not valid in general when more than one measure of event size is used. They also suggested a modified model, which introduces the misallocation of resources. Similar findings are seen in \cite{4596203} that the HOT model is not a generic one that can be used for applications related to power systems, but it was worth studying the model as the tool of power system blackout mechanism analysis and countermeasure. It was also shown in \cite{Hohensee2011PowerLB} that the HOT mechanism can also replicate empirical statistics of power outages. It has to be noted that dimensions of the grid are critical for HOT predictions, as it determines the exponent. Calculating the dimensions of power grids is challenging due to their complex topology, but in general the graph dimension of high-voltage networks is considered to be over 2 \cite{odor2022}.
An interesting point raised by \cite{Hohensee2011PowerLB} is that while both the SOC and HOT models can replicate the size distributions of outages, the main difference might be seen in the temporal correlation of the events, with which SOC could manifest a $1/f^\alpha$ noise \cite{jensen19891,kertesz1990}. However, previously studied data still cover a too short time span, restricting one to assert, in typical power systems, if it is SOC or HOT in play or an interplay of both.
 
By studying more outage data from various sources and time spans, this latter argument served as one of the main motivations for our paper, as we revisit the empirical studies carried out in the last two decades, and perform an in-depth comparison with our recent findings.

When reviewing the corresponding literature, one has to pay attention to the adequate use of the terminology, especially when comparing outages, restorations, and blackouts. While the size and duration of blackouts are characteristic properties of the power system under study, outages caused by component failure are typically individual events. As shown by \cite{dobsonekisheva2022} the duration of outages is usually much shorter than that of blackouts and their restorations, and large blackouts, albeit rare, are much more hazardous to society \cite{hines2009,carreras2016}. One also has to differentiate the type of component leading to the outage; repairs for generation equipment and transmission equipment not only differ in their nature but also in their duration. Generation outages are caused by failure of the machinery at a single site, while transmission outages can occur at multiple locations and are also subject to weather.

For blackouts, the first widely analyzed dataset was the probability distribution of unserved energy in relation to North American blackouts between 1984 and 1998 \cite{dobson2007c}, which showed a power-law-like tail \cite{carreras2004,weron2006} with a decay exponent between 1.3 and 2.0. Blackout statistics along the dimension of energy were analyzed for New Zealand \cite{ancell2005} (exponent $\approx$ 1.6) and China \cite{Weng2006} (1.8), while for the dimension of power, data for North America \cite{926277,weron2006} (2.0) and Norway \cite{refId0} (1.7) were analyzed earlier. Disturbance date of Sweden was analyzed in \cite{holmgren2006} (exponent $\approx$ 1.6). Besides these examples of empirical data, various power system models are also used to study SOC processes; the most relevant being the standard blackout model named ORNL-PSerc-Alaska (OPA). The OPA model was also validated for real-world data in \cite{carreras2019validating}.

Those previous data had mainly focused on more noticeable blackout events that usually resulted from cascading failures. From the aspect of outage and restoration times, substantially longer datasets were analyzed recently in \cite{ROSASCASALS2011805} and \cite{kancherla2018}. The first study, comprised of 7 years of pan-European outage data shows power-law fits with exponents of 1.7 and 2.1 for energy and power dimensions, respectively. The second study analyzes 14-year restoration time data of the Bonneville Power Administration, and a power-law fit (1.84) is shown to be the most suitable among other distributions, but empirical data has a bit heavier tail than the fit.

In this paper, the collected data record all types of forced outage events due to individual component failures, be they almost negligible or affecting consumers from vast regions. The natural question is then if heavy tails could still emerge for accessible statistics, such as for the distributions of unavailable energy and unavailable duration \footnote{It has to be noted that most data providers do not distinguish between outage and restoration processes in their openly published datasets, thus from herein we will refer to outage duration as the time passing between the de-energization and the re-energization of a power system component.}. And if so, what are the implications of these heavy tails? In what follows, we should be poised to answer these questions.
We show by spectral analysis, that the inter-event times and the outage frequencies in the database exhibit a constant and a PL decaying component. The former, valid for short times indicates random events excluding SOC, while the latter, valid for longer time scales suggests an SOC mechanism, in which the optimization of repairing capacity plays 
the role of tuning to a critical point. Note, that even a large constant can suppress 
a PL decay \footnote{The spectral analysis of the outage duration times is less clear. For certain databases 
one cannot see a crossover between random and PL correlations, if the temporal resolution is low.}.
Based on the SOC, we propose a simple model, which can describe the outage/repair competition, taking into account branching-like failure cascades, and which is capable to show outage duration critical exponents close to the observations. This model has already been explored in
the statistical physics of sand piles and reaction-diffusion systems. Based on this model we try to propose possible mechanisms for mitigating outage duration
exponents. However, the spectral analyses of the outage durations tend to suggest that the HOT mechanism, based on maintenance
resource optimization, may be more favorable for describing the empirical outage duration data.

The remainder of the paper is organized as follows. Section \ref{sec:2}. presents the methods and data used for the research. Results for the empirical data are detailed in Section \ref{sec:3}. Discussion and explanation of the results are presented in Section \ref{sec:4}.

\section{Data and methods \label{sec:2}}
\subsection{Outage data description\label{sec:2A}}

We collected outage data from various online public sources of Transmission System Operators (TSOs). From these data, we exclusively consider outages caused by individual component failures, and no blackout events are considered. The interested outage events may be caused by any component failure before the power is transmitted to the customer end, be it a generator failure or any failure in the course of power transmission, the latter of which includes failures such as transmission line failures, circuit breaker failures, switch failures, transformer failures, etc. Hence, the collected data are categorized in accordance with their nature, as follows.

(i) Generator outages. This category includes the European Network of Transmission System Operators for Electricity (ENTSO-E) online data for the unavailability of production and generation units \cite{entsoeprodgen} for control areas (see the definitions of these terms in Appendix~\ref{appenda}), the California Independent System Operator (CAISO) generation outage data \cite{caliiso}, the Hungarian Transmission System Operator (Magyar Villamosenergia-ipari \'{A}tviteli Rendszerir\'{a}ny\'{i}t\'{o}, MAVIR) data on generation outages \cite{huMAVIR}, and the production outage data from the Italian Power Exchange (Gestore dei mercati energetici, GME) Inside Information Platform \cite{gmeitaly}. 

(ii) Transmission outages. As remarked above, many different types of transmission infrastructures can go wrong. For simplicity, we identify two types of transmission outages, i.e.~transmission line outages, which includes not only line failures but also failures in circuit breaker, switches, etc., and transformer outages. In this regard, the Bonneville Power Administration (BPA) data provides both transmission line outages and transformer outages \cite{BPAout}, while the Alberta Electric System Operator (AESO) historical transmission outage data \cite{aesohist} recorded a hybrid of transmission line outages and transformer outages. A special case comes from the ENTSO-E data for the unavailability of transmission grids between control areas \cite{entsotransm}), which only records transmission line outages and transformer outages between pairs of control areas. The data indicate that the majority of outages were caused by transmission line failures though.

Most of these data sources do not provide dataset dumps but rather data access APIs. Hence, we have employed automatic data-scraping methods to assemble as much data as possible for further analyses. Note that automatic data collection methods are particularly useful for data sources that are limited by single-time query quota restrictions. For a detailed account of data collection, see the descriptions in the respective references \cite{entsotransm,caliiso,aesohist,BPAout,huMAVIR,gmeitaly} and the raw data repository \cite{datarepo}.

All the data are formatted in terms of many outage entries containing several columns that detail the information of an outage event, such as the ID of the plant/line/transformer, the starting time and the ending time, the outage type, the unavailable power (or the installed power and the available power), etc. In this paper, we only focus on the statistics of unexpected outages that TSOs usually do not have precautions for prevention, which are usually due to natural causes, such as storms and lightning strikes, or failures of generator components and transmission facilities. To this end, we filtered out entries for any planned outages for maintenance. This amounts to selecting the entries labeled with ``Unplanned outage'' (or ``Forced outage'') for the ENTSO-E data, the entries labeled with the ``FORCED'' outage type for the CAISO data, the entries labeled with the ``Auto'' outage type for the BPA data, and the entries labeled with "UNPLANNED" unavailability type for the Italy GME data, while the Hungary MAVIR data are already for  unplanned outages. The AESO data does not contain such a field for filtering, and we simply selected entries designated with ``Significant Outage''.

Previously, it has been shown that the sizes of blackouts and outages, measured by various quantities (energy, power, affected customers, etc.)  follow power laws characterized by exponents ranging from $\sim1.3$ to $\sim2.0$ \cite{carreras2000,dobson2007c}. However, as remarked in the introduction, although several possible explanations, for example, SOC for the direct current (DC) model \cite{dobson2007c}, cascade failures in the second-order Kuramoto model for alternating currents \cite{odor2020,odor2022}, and power-law city sizes \cite{nesti2020}, have been proposed to understand the emergence of these power laws, a conclusive, full account for the origin of these power laws is not yet reached in the literature. Especially, due to the complexity of power-grid systems and their interaction with the environment, with our everyday life and production activities, and with the plant operation and maintenance staff, then, natural events, population distributions, power-grid infrastructure failures, human societal behaviors, responses of the maintenance staff to outage events, and so on could all contribute to the heavy tails of various measures to a certain extent. In a forthcoming study, we investigate the occurrences of PL-s in the generator capacity sizes~\cite{heterogeneity}. 
As for HOT, power spectra analysis has not been carried out before, according to our knowledge.

To further probe the possible origins of these power laws, we first reproduce similar power-law observations with the collected outage data, with outage sizes measured by the unavailable duration ($T_u$) and the unavailable energy ($E_u=T_u\times P_u$, with the notation $P_u=\text{Unavailable Power}$); see Appendix \ref{appenda} for a full account of how these quantities are obtained from each raw dataset. As will be shown later, empirical unavailable duration distributions could also manifest power-law tails. Since many outage quantities can take the form of $T_u \times \text{certain measure}$, we hence speculate that the understanding of the manifested power law for the unavailable duration constitutes a crucial ingredient for the understanding of general power-law behavior in outage distributions.

\subsection{Methods of outage data analysis\label{sec:2B}}

Suppose the studied measure of outage sizes is denoted by $x$, representing $E_u$ or $T_u$. One can assume that $x$ is drawn from a \textit{continuous} probability distribution $\mathrm{Prob}(x)$. In particular, empirical data usually display a power-law tail starting from a certain minimum value $x_\mathrm{min}$ \cite{clauset2009} \footnote{From Eq.~\eqref{eqs:plpdf}, the probability density diverges as $x\to 0$, so a lower bound is also mandatory mathematically.}. One can further assume the tail part of the distribution is drawn from a \textit{continuous} power-law distribution with the probability distribution function (PDF)
\begin{equation}
    p(x)=\frac{\tau-1}{x_\mathrm{min}}\left(\frac{x}{x_\mathrm{min}}\right)^{-\tau}\,,
    \label{eqs:plpdf}
\end{equation}
and the cumulative probability function (CDF)
\begin{equation}
    P(x)=\int_x^\infty p(x')\mathrm{d}\!x' = \left(\frac{x}{x_\mathrm{min}}\right)^{-\tau+1}\,.
    \label{eqs:plcdf}
\end{equation}
In what follows, when concrete data are studied, we will associate $\tau_E$ and $\tau_T$ with the distributions of $E_u$ and $T_u$, respectively, for distinction.

For an outage dataset of $N$ entries, assuming the power law starts to hold from $x_\mathrm{min}$ so that the remaining $n$ entries whose $x \ge x_\mathrm{min}$ fit into a power law, the exponent $\tau$ can then be estimated by maximizing the logarithmic value of the likelihood
\begin{equation}
    p(x|\tau)=\prod_{i=1}^{n} \frac{\tau-1}{x_\mathrm{min}} \left(\frac{x_i}{x_\mathrm{min}}\right)\,, \quad \forall x_i\ge x_\mathrm{min}\,,
\end{equation}
giving rise to the estimated exponent \cite{clauset2009}
\begin{equation}
    \hat{\tau}=1+ n \left[\sum_{i=1}^n \ln \frac{x_i}{x_\mathrm{min}}\right]\,.
    \label{eqs:tau}
\end{equation}
However, since $x_\mathrm{min}$ is usually not known \textit{a priori}, one can select any  value $x=X$ as $x_\mathrm{min}$ and obtain an estimation $\hat{\tau}(x_\mathrm{min}=X)$. In order to find the optimal estimation $\hat{x}_\mathrm{min}$, we resort to minimizing the Kolmogorov-Smirnov statistic distance \cite{clauset2009}
\begin{equation}
    D(x_\mathrm{min})=\max_{x\ge x_\mathrm{min}}|S(x)-P(x)|\,
\end{equation}
that quantifies the distance between the empirical CDF $S(x)$ of the data for observations $x\ge x_\mathrm{min}$ and the CDF Eq.~\eqref{eqs:plcdf} for the best power-law fit of the data [with $\tau=\hat{\tau}$ estimated by Eq.~\eqref{eqs:tau}] in the region $x\ge x_\mathrm{min}$.

For the unavailable duration $T_u$, to explore the crossover from short-time behavior due to quick routine maintenance to power-law behavior in the much longer maintenance regime, we separate the events with respect to a temporal threshold and perform power-law fittings for $T_u\le 24\ \text{hours}$ and $T_u> 24\ \text{hours}$, respectively. The threshold was determined based on consultations with experts from the fields of power system maintenance and reliability of electric machinery. According to operational experience, forced outages rarely end in permanent failures, as they are caused by random events and the majority of such failures can be corrected in a range of hours. On the other hand, major faults typically require specialized equipment and/or personnel, which is available to a limited extent on-site. Service level agreements for operation and maintenance typically consider repairs to be started within 24 hours, but as the correction of these major faults usually requires mechanical work and possible disassembly of the equipment, registered outage times are exceeding 24 hours.

\subsection{Methods of spectral data analysis\label{sec:2C}}
For major outage events, where large-scale blackout could occur, the entire outage duration distribution (and its power-law tail) may be accounted for by the inverse Weibull distribution ($\alpha x^{-\beta-1}e^{-\lambda x^{-\beta}}\sim \alpha x^{-\beta-1}$ for $x\gg 1$) due to the symmetry of failures and restorations \cite{harvey2004}. However, it is also possible that the studied data may include many intermittently occurred random outage events, then contrary to Ref.~\cite{Wu2022}, the inverse Weibull distribution could not fit the data to the whole range.

To demonstrate that the studied outage data are indeed comprised of random outage events, characterized by white noise signals, as well as correlated events potentially resulting from cascading outages, we perform power spectral analyses \cite{jensen19891,kertesz1990} on the time series $S(t)$, which, similar to the number of ``topplings'' in the sandpile model \cite{bak1987}, is the number of outage events at time $t$, and on the time series $I(t)$ for the $t$th interval between successive outage events. In addition, the duration $D(t)$ for the $t$th outage event can also be regarded as a time series and should be analyzed. For $S(t)$, we have used the accessible time resolution, either one hour or one minute, as the time unit, while for $I(t)$ and $D(t)$, the time $t$ bears the meaning of the index of an event interval and the index of an event, respectively \footnote{For both $S(t)$ and $I(t)$, we had first excluded events recorded at 1/4, 1/2, 3/4 hours and whole hours to eliminate any artifacts in outage bookkeeping time, if many events were recorded at such time points.}. As an intuitive example for showing the effect of white noise on correlated signals, we also performed power spectral analyses on the noise signal $N(t)$ obtained by superposing a Brownian noise with white noise of different intensities
\begin{equation}
        N(t)=\sum_{i=1}^{t} X_i + Y\,, \quad t=1, 2, \dots, T\,,
\end{equation}
where $X_i$ and $Y$ are random variables drawn from the normal distributions $\mathcal{N}(0,1)$ and $\mathcal{N}(0,\sigma)$, respectively, with $\sigma=0, 1,$ and $10$. The power spectrum $P_{\mathcal{F}}(f)$ of a signal $\mathcal{F}(t)$ [$=S(t)$, $I(t)$, $D(t)$, or $N(t)$] is then computed as the absolute square of the discrete Fourier transform of $\mathcal{F}(t)$ (via the fast Fourier transform) \cite{brown2012}
\begin{IEEEeqnarray}{rCl}
        H(f&=&\frac{t}{T}) = \frac{1}{\sqrt{T}}\sum_{k=1}^{T}\mathcal{F}(k) e^{2\pi i (k-1) (t-1)/T}\,,\\
        P_{\mathcal{F}}(f) &=& |H(f)|^2+|H(1-f)|^2\,, \,\,0 < f\le \frac{1}{2}\,.
\end{IEEEeqnarray}
We remark that this analysis directly probes the correlation of raw outage events as well as their restorations, it is in contrast to probing the correlation of blackout events through their long-term correlations and the waiting times between successive blackouts \cite{carreras2016}.

\begin{figure*}[!htbp]
        \centering
        \includegraphics[width=0.99\textwidth]{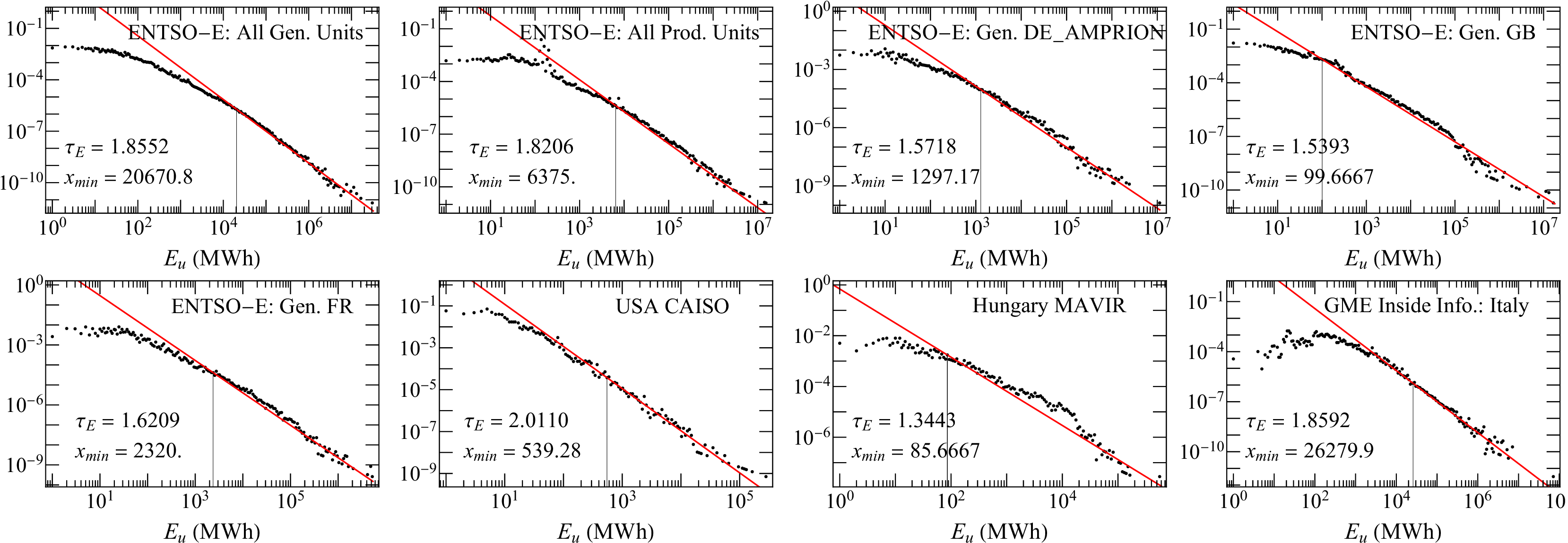}
        \caption{Probability distributions (black dots) of generation outages measured in terms of the unavailable energy. For the ENTSO-E data, we show the generation outage data for the control areas ``DE\_AMPRION'', ``GB'', and ``FR'', as well as the generation and production outage data from all control areas. The fitted power laws and their corresponding $x_\mathrm{min}$ values are marked by solid red lines and vertical black lines, respectively.}
        \label{fig:alltsoen}
\end{figure*}

\begin{figure*}[!htbp]
        \centering
        \includegraphics[width=0.99\textwidth]{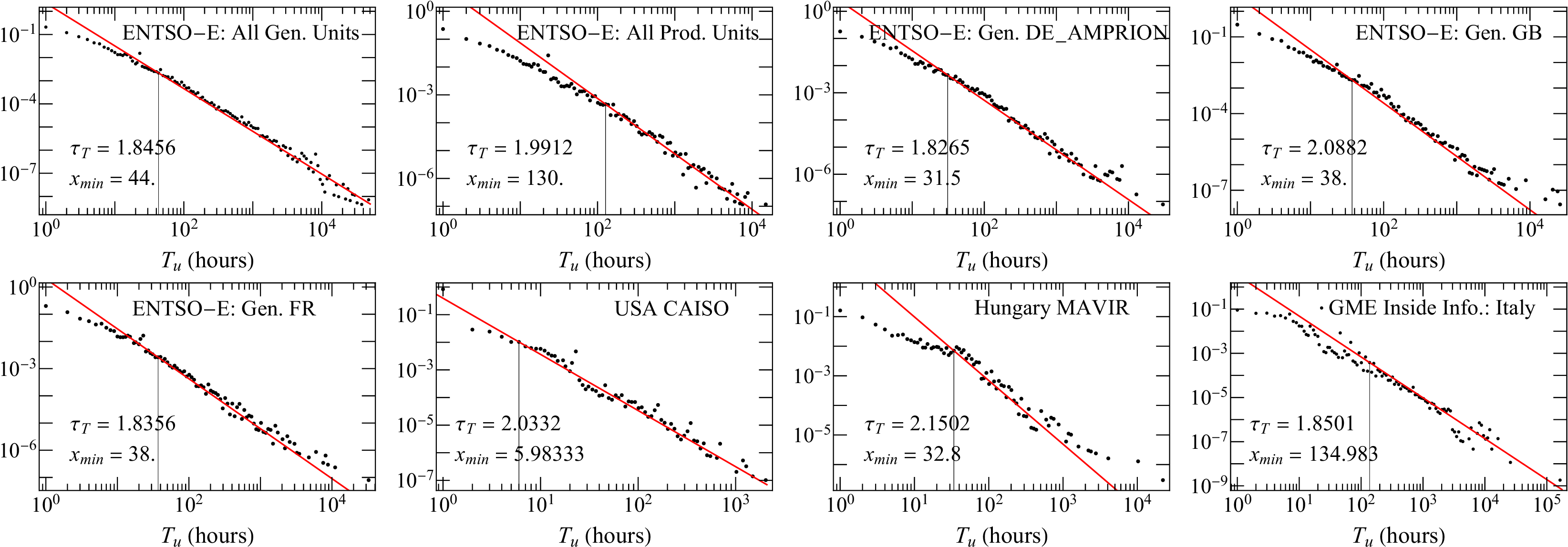}
        \caption{Probability distributions (black dots) of generation outages measured in terms of the unavailable duration. For the ENTSO-E data, we show the generation outage data for the control areas ``DE\_AMPRION'', ``GB'', and ``FR'', as well as the generation and production outage data from all control areas. The fitted power laws and their corresponding $x_\mathrm{min}$ values are marked by solid red lines and vertical black lines, respectively.}
        \label{fig:alltsodu}
\end{figure*}

\section{Results \label{sec:3}}

We categorize the outage statistics into outages in generation/production units and outages in transmissions (see Appendix~\ref{appenda} for the difference between generation and production units). Note that transmission outages may be caused by various transmission infrastructure failures in transmission lines, transformers, circuit breakers, switches, etc.

\subsection{Results on generator outages\label{sec:3A}}
Figs.~\ref{fig:alltsoen} and \ref{fig:alltsodu} show the probability distributions of outages in generation units and production units from various data sources as described in Sec.~\ref{sec:2A}, measured in terms of unavailable energy and unavailable duration, respectively. In our analysis we considered both deratings of generation units and shutdowns; in the first case, the output capacity of the generator is lower than its nominal value, while in the latter case, we assume the complete loss of power output. In both cases, the volume of unavailable energy was determined by multiplying the duration of the outage with the power output considered to be unavailable. In real operating conditions, generator outages may present a wider spectrum, as presented in \cite{MURPHY20181360,MURPHY2019113513}, where the authors also highlight that while generator outages are widely considered as individual events, significant correlations can be discovered between the outages.

\begin{figure*}[!htbb]
    \centering
    \includegraphics[width=0.99\textwidth]{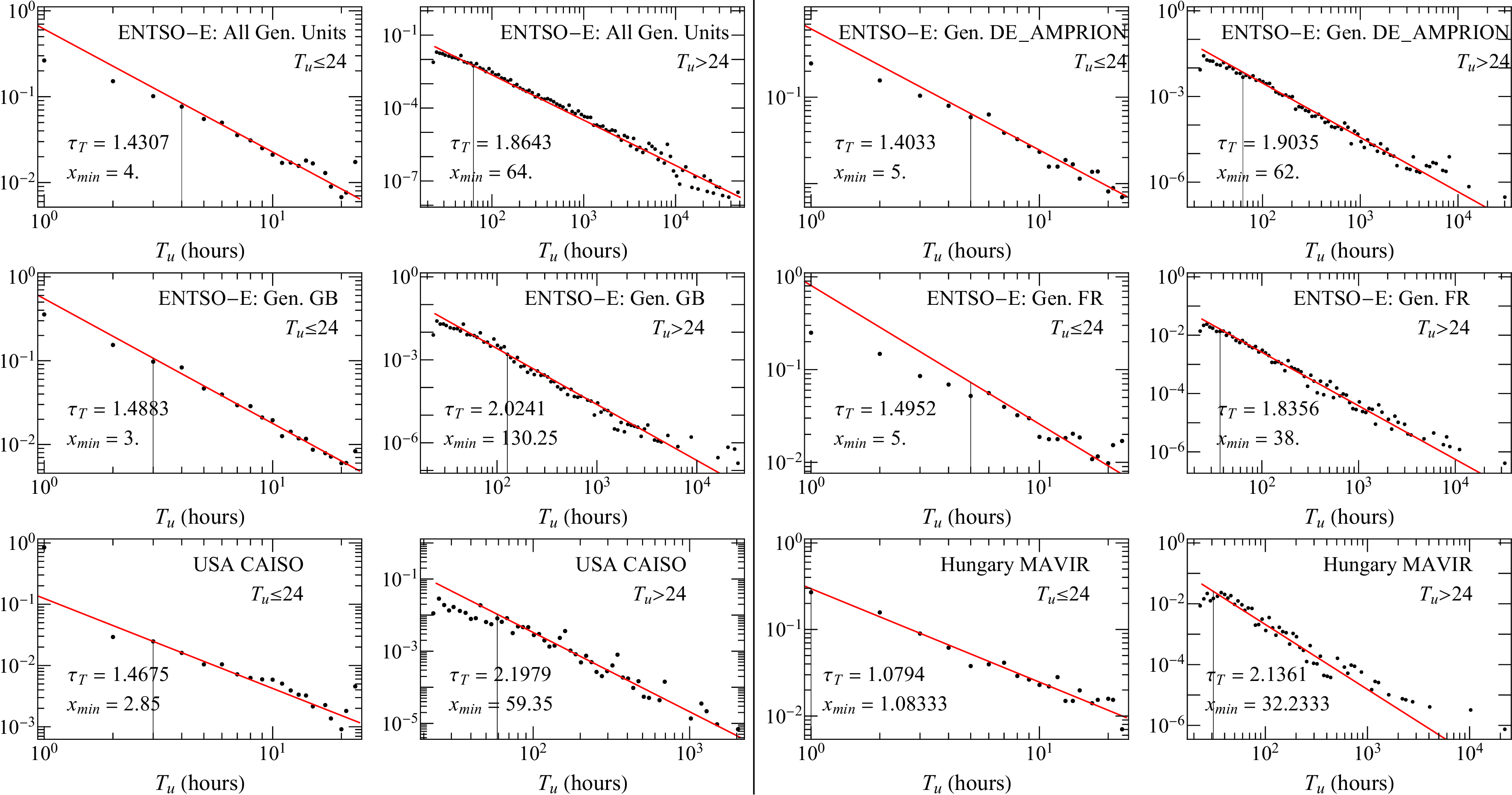}
    \caption{Probability distributions (black dots) of generation outages measured in terms of the unavailable duration, with the duration separated by a threshold at $24$ hours. The fitted power laws and their corresponding $x_\mathrm{min}$ values are marked by solid red lines and vertical black lines, respectively.}
    \label{fig:duth24}
\end{figure*}

For the ENTSO-E data, we show results for all generation units and all production units, as well as results for several large control areas, including ``DE\_AMPRION'', ``GB'', and ``FR'' \cite{Note2}. The distributions were obtained by binning the data logarithmically with base 1.08 first and then plotted in a log-log scale.  These data distributions clearly display power-law tails. We employed the fitting method detailed in Sec.~\ref{sec:2B} to determine $x_\mathrm{min}$ and the power-law exponent $\tau_E$ or $\tau_T$; also c.f.~Table \ref{tab:exps} for a summary of exponent values. For generation/production outages measured by the unavailable energy, $\tau_E$ ranges from $\sim1.3$ to $\sim2.0$, and for generation/production outages measured in terms of the unavailable duration, $\tau_T$ ranges from $\sim1.8$ to $\sim2.1$. In addition to the statistics for generation/production units, Fig.~\ref{fig:alltsodu} also shows the unavailable duration distributions for BPA customer services and BPA transformers, with $\tau_T\simeq 1.83$ and $\tau_T\simeq 1.09$, respectively. Although here we only study single outage events, the observations for energy outage are thus comparable with the literature for blackout events \cite{carreras2000,carreras2004,weron2006,holmgren2006,Weng2006,ROSASCASALS2011805,dobson2007c}. What is more, the listed outage duration distributions further exemplify the findings for the transmission line restoration duration in Ref.~\cite{kancherla2018} (there a power-law tail was identified with $\tau_T\simeq 1.84$), however, here for the restoration of outages in both generation and transmission facilities; see also the next subsection.

As remarked in Sec.~\ref{sec:2B}, the 24-hour threshold typically marks a separation for short- and long-time maintenance behaviors. One may then expect to observe quite different statistics for $T_u\le 24$ hours and for $T_u > 24$ hours. This is immediately justified by Fig.~\ref{fig:duth24}, upon applying the 24-hour threshold on the duration of generation outages. Both the distributions for $T_u\le 24$ hours and $T_u > 24$ hours display power-law tails (although distributions for the former case give weaker power laws with few orders of magnitude due to the threshold), hinting that maintenance activities are governed by different universal behaviors in short- and long-time maintenance.

\begin{table*}[!htbp]
\begin{center}
        \caption{Summary of various exponents obtained for energy outages ($\tau_E$) and for outage duration ($\tau_T$), with available $\tau_T$ for $T_u\le 24$ hours (denoted as $\tau_{T_{\le 24}}$) and $T_u>24$ hours (denoted as $\tau_{T_{> 24}}$) also displayed. The lower right part of the table shows results for transmission outage duration between pairs of control areas in the ENTSO-E data. Note that a similar set of exponents exist for pairs of control areas in a reversed transmission order (such as ``ENTSO-E SE-NO''). The corresponding exponents are quite close to the ones shown in the lower right part of the table and are not included.}\label{tab:exps}
        \begin{tabular}{lllllllll}
                \toprule
                Generation & $\,\tau_E\,$ & $\,\tau_T\,$ & $\,\tau_{T_{\leqslant 24}}\,$ & $\,\tau_{T_{> 24}}\,$ & \hspace{0.8em} & Transmission & $\,\tau_T\,$ & $\,\tau_{T_{> 24}}\,$ \\
                \hline
                ENTSO-E All Gen. & 1.86  & 1.85  & 1.43 & 1.86  & &  BPA Transmission & 1.85 & 1.72 \\
                ENTSO-E All Prod. & 1.82  & 1.99  & 1.34 & 2.24  & & BPA Transformer & 1.09 & 1.17 \\
                ENTSO-E DE\_AMPRION $\,\,\,$ & 1.57 & 1.83 & 1.40 & 1.90 & & AESO & 2.37 & 2.37 \\
                ENTSO-E GB & 1.54 & 2.09 & 1.49 & 2.02 & & ENTSO-E All Transmission & 1.54  & 1.54 \\
                \cline{7-9}
                ENTSO-E FR & 1.62 & 1.84 & 1.50 & 1.84 & & ENTSO-E DE\_50HZ-PL\_CZ $\,\,\,$ & 1.07 & 1.01 \\
                USA CAISO & 2.01 & 2.03 &  1.47 & 2.20 & & ENTSO-E NO-SE & 1.12 & 1.12 \\
                Hungary MAVIR & 1.34 & 2.15 & 1.08 & 2.14 & & ENTSO-E PT-ES & 1.25 & 1.25 \\
                GME Italy & 1.86 & 1.85 & 1.54 & 1.85 & & ENTSO-E SE-DK\_CA & 1.00 & 0.97 \\
                \bottomrule
        \end{tabular}
\end{center}
\end{table*}

\subsection{Results on transmission infrastructure outages}

\begin{figure}[!htbp]
    \centering
    \includegraphics[width=0.99\columnwidth]{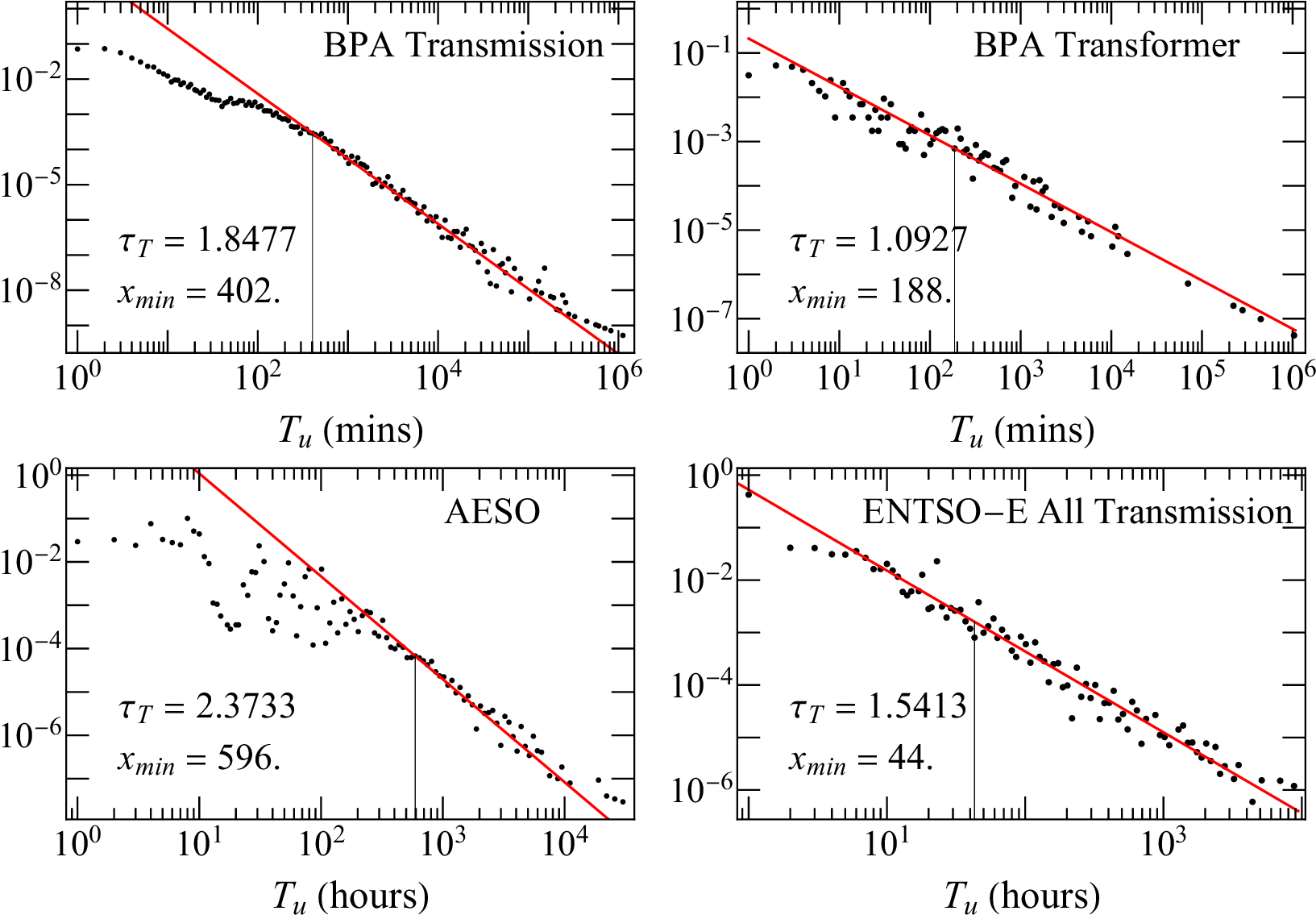}
    \caption{Probability distributions (black dots) of various types of transmission outages measured in terms of the unavailable duration. The fitted power laws and their corresponding $x_\mathrm{min}$ values are marked by solid red lines and vertical black lines, respectively.}
    \label{fig:dutransmts}
\end{figure}

Unlike centralized power plants, transmission infrastructures are much more widely distributed over large regions. As such, in the way of power transmission, any failed transmission lines, switches, circuit breakers, transformers, etc., could cause outages. The BPA data provides both outage duration data for transmission lines (including switch and circuit breaker failures) and outage duration data for transformers separately; the data are shown in the two top panels of Fig.~\ref{fig:dutransmts}. In these two cases, as in Ref.~\cite{kancherla2018}, the power-law tails are quite evident, and the smaller exponent for transformer outages, in accordance with our intuition, suggests that restoring the functionality of a transformer is more likely to take a longer time as compared to repairing a transmission line. Even if the data contain a mixture of outage events caused by both transmission line failures and transformer failures, as shown by the two bottom panels of Fig.~\ref{fig:dutransmts} for the AESO data and the combined ENTSO-E data for transmission outages between all pairs of control areas (see below for a few examples on some selected control area pairs).

\begin{figure*}[!htbp]
    \centering
    \includegraphics[width=0.99\textwidth]{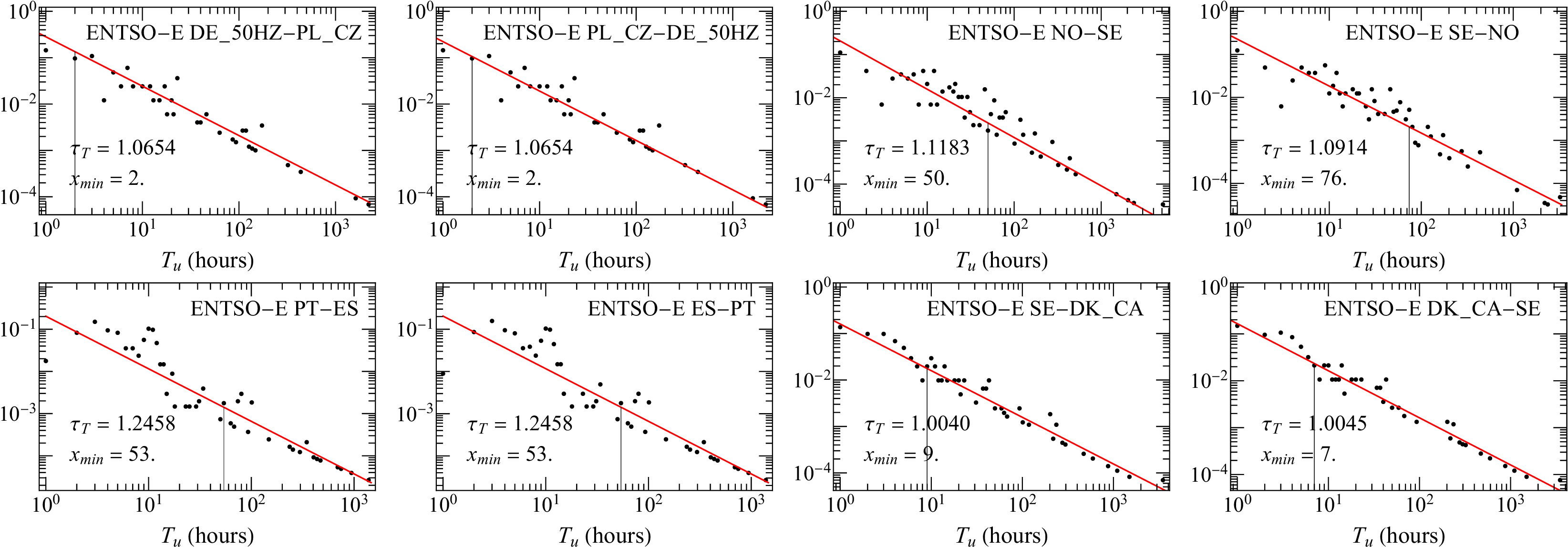}
    \caption{Probability distributions (black dots) of transmission outages between  pairs of control areas (labeled on the top of each panel), measured in terms of the unavailable duration. The fitted power laws and their corresponding $x_\mathrm{min}$ values are marked by solid red lines and vertical black lines, respectively.}
    \label{fig:ducontrol}
\end{figure*}

\begin{figure*}[!htbp]
        \centering
        \includegraphics[width=0.99\textwidth]{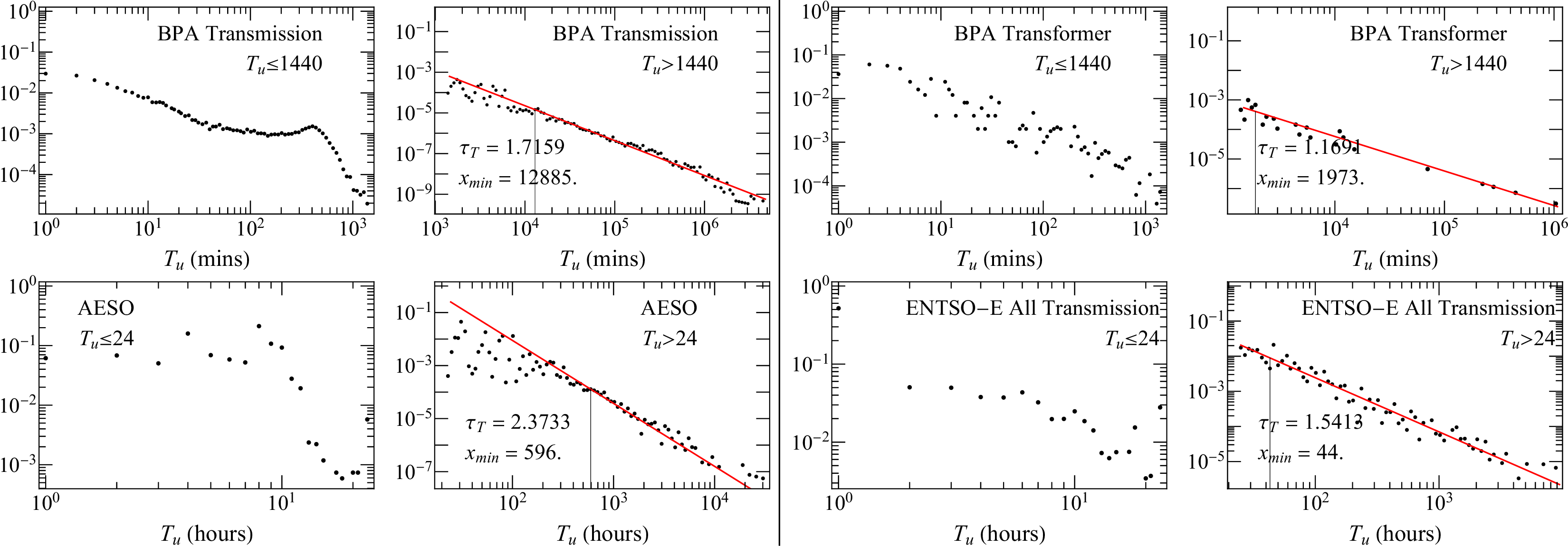}
        \caption{Probability distributions (black dots) of transmission outages measured in terms of the unavailable duration, with the duration separated by a threshold at $24$ hours (1440 minutes). For $T_u>24$ hours, the fitted power laws and their corresponding $x_\mathrm{min}$ values are marked by solid red lines and vertical black lines, respectively. For $T_u\leq 24$, unlike in Fig.~\ref{fig:duth24}, the statistics are not justifiable for power-law fits.}
        \label{fig:dutransth24}
\end{figure*}

\begin{figure*}[!htbp]
        \centering
        \includegraphics[width=0.99\textwidth]{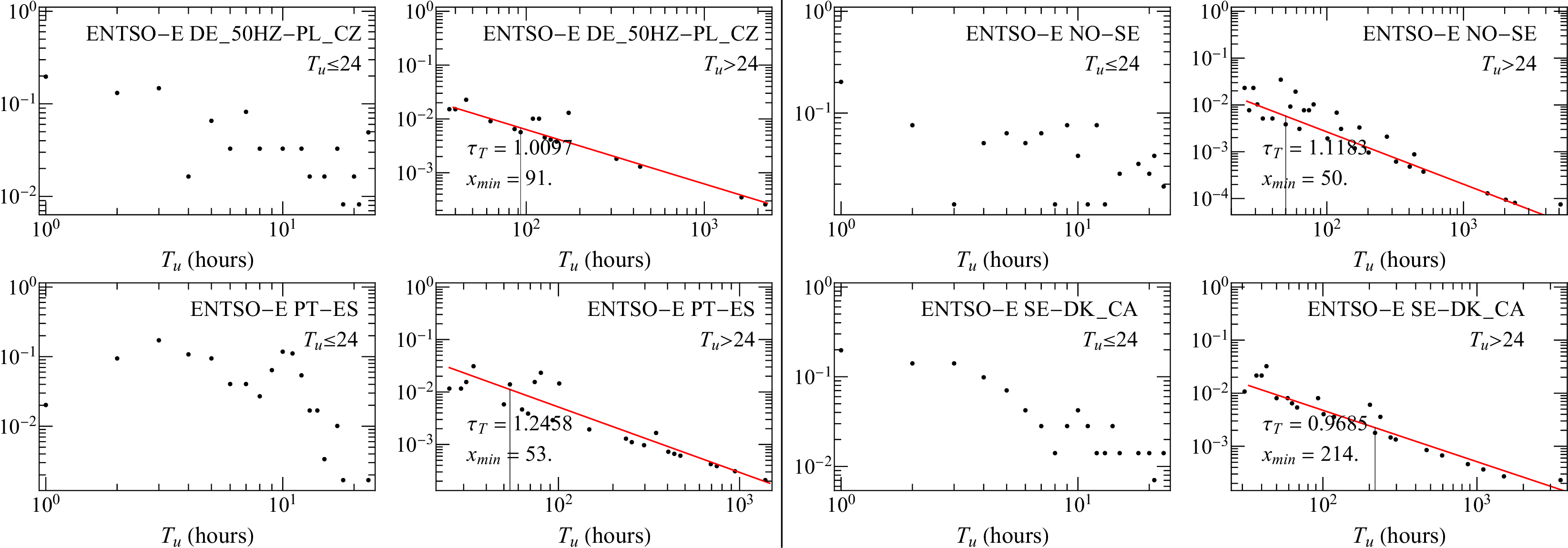}
        \caption{Probability distributions (black dots) of transmission outages measured in terms of the unavailable duration, with the duration separated by a threshold at $24$ hours (1440 minutes). For $T_u>24$ hours, the fitted power laws and their corresponding $x_\mathrm{min}$ values are marked by solid red lines and vertical black lines, respectively. For $T_u\leq 24$, unlike in Fig.~\ref{fig:duth24}, the statistics are not justifiable for power-law fits.}
        \label{fig:dutransthcontrol24}
\end{figure*}

Hence, not only transmission line failures and transformer failures but also their mixture could lead to power-law outage distributions. The latter can be made sensible via a mixture distribution \cite{mclachlan2019}. For a data source with $x$ drawn from several distributions $p_1(x), p_2(x), \dots, p_n(x)$ and weights $w_1, w_2, \dots, w_n$, where $w_i>0$ and $\sum_i^n w_i=1$, the mixture distribution for $x$ would be $p_m(x)=\sum_i^n w_i p_i(x)$. Let us assume that general transmission outage events can be caused by both transmission line failures and transformer failures, which follow the power-law distributions $p_1(x)=k_1 x^{-\tau_{T1}}$ and $p_2(x)=k_2 x^{-\tau_{T2}}$ respectively in the tails. Now if the events happen with weights $w_1$ and $w_2$, the tail part [$x>\left(k_1/(\tau_{T1}-1)\right)^{\frac{1}{\tau_{T1}-1}}$ and $x>\left(k_2/(\tau_{T2}-1)\right)^{\frac{1}{\tau_{T2}-1}}$] of the mixture distribution will be given by
\begin{IEEEeqnarray}{rCl}
    p_m(x)&=&\frac{1}{w_1 + 
    w_2} \left(w_1 k_1 x^{-\tau_{T1}}+w_2 k_2 x^{-\tau_{T2}}\right)\nonumber \\
    &\sim &x^{-\tau_T}\,, \label{eqs:mix}
\end{IEEEeqnarray}
so that the tail is dominated by the power law of the smaller exponent $\tau_T=\min(\tau_{T1},\tau_{T2})$, as long as $w_1\sim w_2$.

The above argument can be now readily extended to understand the observed power law of the ENTSO-E transmission outage duration for all pairs of control areas, given each pair of them follows a power law. To check this, in Fig.~\ref{fig:ducontrol} we show transmission outage duration distributions for a few pairs of control areas. Since the sizes of most such datasets are too small to give good enough statistics, the displayed data are specifically selected for pairs of larger control areas that contain enough outage events. From the figure, we see that for these larger control areas, the transmission outages between them indeed follow power laws in outage duration distributions and typically have an exponent $\tau_T$ ranging from $\sim 1.0$ to $\sim 1.3$, suggesting that transmission infrastructures between larger control areas are more difficult to be fixed if they fail. Since the exponents in these examples are fairly small, then according to \eqref{eqs:mix}, it is tempting to conjecture that the overall exponent will be $\tau_T\sim 1$ if we combine the data for all pairs of control areas. However, out of 226 datasets for different pairs of control areas, only a few are large enough to show meaningful statistics, so their weights are almost negligible. Hence, the exponent $\tau_T\sim 1.54$ from Fig.~\ref{fig:ducontrol} must be ascribed to the overall statistics of smaller control areas and theoretically, they should be governed by larger exponent values, even though the dataset of each of them is yet too small to show significant statistics.

In Figs.~\ref{fig:dutransth24} and \ref{fig:dutransthcontrol24}, a 24-hour threshold is again applied to the duration of transmission outages corresponding to Figs.~\ref{fig:dutransmts} and \ref{fig:ducontrol}, respectively. In contrast to generation outages in which a 24-hour separates two different universal behavior due to maintenance teams' responses in a specific power system, transmission outage duration only show power-law tails for $T_u> 24$, while for $T_u\le 24$ hours, power-law tails are hardly observed due to either a genuine lack of power laws or probably too small sample sizes. If the former is the case, it may be that transmission lines usually locate in remote places, many of which could only be fixed with a longer time so that only longer-time maintenance activities follow a universal behavior.

\section{Discussion \label{sec:4}}
We summarize the obtained exponents in Table \ref{tab:exps}. As mentioned earlier, the exponent $\tau_E$ for the unavailable energy shows good agreement with the literature. For outage duration in the generation sector, we conclude that a 24-hour threshold typically renders $\tau_{T_{\le 24}}\sim 1.0-1.5$ and $\tau_{T_{> 24}}\sim 1.80-2.20 \sim \tau_T>\tau_{T_{\le 24}}$. For transmission infrastructure outage duration, even though the statistics are not good enough for power-law fits when $T_u\le 24$, we still see that $\tau_{T_{> 24}}\sim \tau_T$. Hence, the observed power-law tail for $T_u$ for each dataset is essentially unaffected if outage events with $T_u\le 24$ are all excluded. These results suggest that typical outages not yet restored beyond 24 hours are governed by a universal mechanism that gives rise to heavy tails.

\subsection{Potential explanations for heavy tails}
Since many outage measures are related to the duration of the outage events, these observations for unavailable duration distributions thus provide another perspective to understand the ubiquitous existence of heavy tails in power-grid systems. In the following, potential explanations of the underlying phenomena are presented. The explanations are structured in three sections; (i) discusses potential relation to SOC, (ii) discusses that to HOT, while (iii) is a summary, arguing for more analysis and the need for more detailed empirical data.

(i) Empirical data of faulty electrical (and many other types) components may show exponential distributions~\cite{duffey2019}, but in general had been described by lognormal, gamma or Weibull repair time distributions~\cite{4641852}. Furthermore, the heavy-tailed outage durations we observe in the databases may be the consequence of dependent events. In section \ref{sec:4B} we provide a power-spectral analysis, which shows power-law decaying auto-correlations for quite a proportion of the outage events.
A possible explanation for the correlations among the repairs can be related to the limited capacity or availability of the maintenance staff/equipment within a region. Thus, the observed power laws can have a similar origin as those of the cascading blackout events themselves: self-organized criticality (SOC) \cite{bak1987,bak2013}, tuned by the competition of  supply and demand, to the edge of a critical point \cite{carreras2000,dobson2007c}, which is optimal for the function of the whole economy \cite{stanley2002}. In other words, keeping too large a maintenance capacity is economically inefficient, while too small is dangerous for the function of the whole power grid, therefore the maintenance staff/resources of the system tune them to a SOC state as in the case of power production. One can naturally view the heavy tails in outage duration distributions as a direct result of the responses of maintenance resources to outages of different scales driven by SOC.
    
To understand the critical exponents and how the crossover behavior can emerge, one can map the outage process onto simple non-equilibrium reaction-diffusion (RD) models with phase transitions. Such mappings have been shown to be very efficient to describe universal behaviors of non-equilibrium systems~\cite{odorbook}. Let us quantify outage sizes in terms of the time integral of failed units until repair. SOC appears, when a slow drive (particle adding in RD models) and a faster dissipation (particle removal in RD models) mechanism competes~\cite{dickman1998,dickman2000}.  
Denoting functioning units by $0$ ``particles'', faulty (active) ones by $A$-s, and the repairing teams by $B$-s we can set up the following reaction scheme
\begin{equation}
    0 \to A, \ \ A + 0\to A+A, \ \ A+B \to 0+B \ ,
\label{eqs:react}
\end{equation}
where the number of $B$-s can be a considered conserved quantity. They move in the system and their role is only to sustain the system's function. 
The $0 \to A$ describes the slow drive, while the $A + 0\to A+A$ models the fast redistribution ending by the dissipative $A+B \to 0+B$ repair process, in agreement
with an SOC model. The outage size is then measured by the time integral of the number of $A$-s in an SOC avalanche.
This simple system can be considered as the combination of spontaneous isotropic percolation~\cite{IP} (IP) and a branching process, describing a possible failure cascade, which implies the criticality of the directed percolation~\cite{DP}, coupled with a conserved density (DP-C)~\cite{odorbook}, as $B$ can be regarded as a background and the total number $N_A+N_0$ is conserved~\cite{van1998}. The process takes place on a network of interacting electrical units which may contain long-range interactions. The interacting network can be quite different from the physical connection network, which is presumably small-world like, and for high-voltage nodes, we showed \cite{odor2020,odor2022} that power grids have graph dimensions between 2 and 3, but we don't know the network of the components of the database. For example, it has been shown in Ref.~\cite{zhou2020} that cascading transmission line outages can be described by Markov chain transition matrices, with which non-local interactions of transmission lines can be observed. When there is no particle number conservation in the system, we know that for regular networks the DP critical fixed point is stable asymptotically as compared to the IP \cite{IP-DP-cross-FT,IP-DP-cross}, so a crossover may happen between IP to DP. 
Such crossover between critical points has also been studied in the IP + DP model in the case of brain models~\cite{PRX-Kor}, on different complex networks. 
Then similarly, with the conserved reactions \eqref{eqs:react}, we should expect an IP to DP-C crossover.
If the spontaneous failure probability is small, it does not cause a critical DP-C percolation, but a slightly off-critical one, which cannot be distinguished from wandering around DP-C criticality as in the case of SOC-like models \footnote{Numerical analysis showed, that in the presence of such composite reactions, the cascade size and duration distribution exponents crossover to larger values on complex networks~\cite{IP-small,IP-PL} for larger avalanche sizes and durations. This agrees and may explain our power-law fitting results for the failure durations, where we see a crossover around $t_c \simeq 24$ hours from smaller to bigger $\tau_T$ exponents. Note, that in the databases the outage times are rounded to 1-min or 1-hour slots, so shorter-time results may not be reliable}.

In case of this DP-C SOC process, we can expect the occurrence of avalanches, triggered by failures of single units, with a power-law size distribution. As for the outage duration distribution, the outage times of single units can be related to the auto-correlation function, which gives the probability of an outage $A$ still not yet being restored after $T_u$, and which exhibits the asymptotic scaling: $C_{AA}(T_u)\propto T_u^{-\lambda/Z}$, where $\lambda$ is the auto-correlation exponent, and $Z$ is the dynamical exponent of the critical process~\cite{odorbook}. Put in another way, this probability can as well be expressed in terms of the outage duration distribution $P_T(T_u)$ as follows
\begin{equation}
    \int_{T_u}^\infty P_T(T) \mathrm{d}T=C_{AA}(T_u)\,,
\end{equation}
giving rise to $P_T(T_u)\sim T_u^{-\tau_T}=T_u^{-\lambda/Z-1}$.

Since the inert 0-s are intrinsically not moving, the scaling properties of reactions \eqref{eqs:react} then belongs to the so-called Manna universality class \cite{odorbook,henkel2008}, which also encompasses the conserved threshold transfer process (CTTP) in $d\ge 2$ dimensions \cite{lubeck2003,odorbook,henkel2008}. By utilizing the scaling relation $\lambda=d-\Theta Z$ \cite{odorbook,henkel2010}, where $\Theta$ is the initial slipping exponent, we obtain
\begin{equation}
    \tau_T=d/Z-\Theta+1\,.
\end{equation}
Inserting the numerical values for $\Theta$ and $Z$ from Ref.~\cite{odorbook,henkel2008} gives $\tau_T\approx 1.37, 1.99$, and $2.51$ in 1, 2, and 3 dimensions. The exponent value in 2 dimensions seems to agree with our database analysis for generation outages that have resulted in $1.8 \le \tau_T \le 2.1$, depending on the region of the data. The exponent values for transmission outages span a wider range, but still seem to be close to the model values in 1 to 3 dimensions. These exponents suggest a fairly universal asymptotic behavior. Differences from the theoretical values can be the consequence of SOC quasi-critical behavior by the spontaneous failures, under-sampling, different networks with different dimensions, or even Griffiths effects ~\cite{griffiths1969,munoz2010}, which occur in case of quasi-static heterogeneity of the system.
    
We don't have data for the failure cluster size distributions, but for the distribution of lost energy $E_u$, which is related to the product of outage duration $T_u$ and the unavailable power capacity $P_u$ of the nodes, which also shows power-law distributions for certain databases (for example in the CAISO data), the power-law tails of $E_u$ follow immediately from that of $T_u$.

(ii) The above-simplified model unavoidably leaves out many complications. First, as outages do not necessarily occur in a SOC cascading manner, the heavy tails in outage duration distributions may not be fully accounted for by the responses of the limited maintenance resources to the SOC cascading failures, as we assumed above. Second, the \textit{ad hoc} $B$ agents themselves, albeit being considered conserved for simplicity, should be organized as a result of further hidden mechanisms for achieving economically efficient responses to outages, so they are not strictly conserved in a long-time span. What is more, the above simple model only regards $B$-s as a background so that their spatial distribution as well as how they will move in responses to the generations of $A$-s are totally discarded, whereas, in reality, some generators in the network may be considered more crucial than others or have to provide higher availability or meet contractual maximum outage criteria. Operators keep extra resources for such facilities to recover more quickly from outage types which would normally cause a longer outage time, thus skewing the outage time distribution towards smaller outage times.

Since man-made complex systems like power-grid systems are highly structured and dominated by optimization designs, the above considerations point to another mechanism for the observed heavy tails: the HOT mechanism \cite{PhysRevE.60.1412,doyle2000}. For forest fires and sandpiles as exemplified in Ref.~\cite{PhysRevE.60.1412}, the HOT mechanism introduces barriers (the resources) to minimize the expected size of the event. These barriers are concentrated in the regions which are expected to be most vulnerable, leaving open the possibility of large events in less probable zones. So in this way, power-law size distribution can emerge by minimizing the average event size via the variational principle with respect to the resource distribution function, subjecting to the restriction of the resources which is used to construct the Lagrange multiplier. \mycomment{\red{delete from here->}Similar barriers are indeed introduced in power systems. The size of an outage is physically limited both on high-voltage and medium-voltage networks. As high-voltage networks tend to be looped, the removal of a line due to an outage should not affect consumers. In the case of medium voltage lines, the radial topology is constructed in an ``arc-ring'' topology, that in case of a fault, circuit-breakers and switches are operated to separate the location and to minimize the number of affected consumers.\red{<-to here}}Hence, even for outage events, although the SOC mechanism can play a role regionally, we cannot exclude that, in pursuit of optimization in maintenance resources distribution and minimization of outage times, the HOT mechanism could partially contribute to the heavy tails in outage size distributions globally.

To be more specific, consider that \textit{independent} cascading outage events, be they uncorrelated single outage events or blackout events, can be indexed by $1\le i\le N$. \mycomment{\red{(Such assumption would not be correct in the case of outage leading to blackouts, as cascading events are inherently not independent, but the data analysed in this paper does not include blackouts.)}} In the original HOT mechanism, 
the resources $r_i$ are allocated to suppress an event of size $l_i$ from happening , obeying a map $l_i=F(r_i)$.
By minimizing the expected outage size \cite{doyle2000}
\begin{equation}
J=\left\{\sum p_i l_i | l_i = F(r_i), \quad \sum r_i\le R \right\}\,,
\label{eqs:hotL}
\end{equation}
subject to the limited resource $R$, it was shown that the size distribution follows a power law $P(l)\sim l^{-\alpha}$. In the above expression, $p_i$ denotes the probability of event $i$ during some time span of observation. The general idea is of course to allocate the resources to places where a large-size event is more likely to happen. However, in the context of outage duration, the focus becomes utilizing the limited resources $R$, say the available maintenance manpower in hours, to fix outages last for various times $T_u$ during the observation time span. Now an event should be directly considered as an outage caused by a generator failure or a transmission infrastructure failure. These events of course do not necessarily occur independently, but for a long enough time span, each of these outage events, indexed by the location $i$ for instance, should associate with a probability $p_i$ and an expected restoration time (outage duration) $T_{ui}$. In this sense, the outage events are again considered as \textit{independent}, and by extending the HOT mechanism \eqref{eqs:hotL}, the observed heavy tails in outage duration distributions may be attained by minimizing the following objective cost
\begin{equation}
J=\left\{\sum p_i T_{ui} | T_{ui}=g(r_i), \quad \sum r_i\le R \right\}\,.
\label{eqs:hotT}
\end{equation}

For the above purpose, the maintenance manpowers are naturally concentrated in regions where outages happen more frequently, leaving outage events in remote places needing longer time to fix, even if it is just a trivial one. Hence, heavy tails $P_T(T_u)\sim T_u^{-\tau_T}$ ensue in outage duration distributions in a similar manner as the original HOT for outage sizes. Nevertheless, at this stage, the HOT explanation can only be a hypothesis as we still lack the knowledge of $p_i$ and the functional form of $g(r)$. What is more, while it is critical to know the dimension of the grid for HOT predictions for cascading sizes, for a more general quantity like the outage duration $T_u$, $g(r)$ is determined by the resource/loss relationship, which may or may not be directly related to the physical dimension \cite{doyle2000}.

(iii) Our empirical results show similarities with experiences of the power sector as well. Ref.~\cite{en13112736} concludes based on the data of a regional power supply company, that the most frequent causes for transmission outages are shortcomings in maintenance, natural and weather conditions, and unauthorized personnel. It is also known that preventive maintenance highly affects the availability of the equipment, but it comes at a cost (monetary and human resources) \cite{RePEc:ecl:harjfk:rwp05-027,en13143571}.

By considering all these intertwined factors, we see that there are various aspects for the emerged heavy tails in outage statistics. On the one hand, some intervention measures, which are usually not optimized, are introduced to the infrastructure, so that outage events are not entirely occurring in a spontaneous SOC manner, but SOC processes could still dominate in many parts of the system. On the other hand, in response to outage events, maintenance resources enjoy bigger flexibility and may be optimally deployed as needed. In summary, our above discussion favored a SOC explanation for the occurrences of outage events, but we don't entirely exclude HOT, whereas, for outage duration, it is more plausible to ascribe the observed heavy tails to an extended HOT mechanism. For outages, since the HOT mechanism restricts more probable large-size events from happening, it is suggested that the $1/f$ spectra may not be observed, contrary to SOC processes \cite{Hohensee2011PowerLB}. To partially unravel if it is SOC or HOT in play for outage events and their duration, in the next section, we try to detect if there are any traits of $1/f$ noise by performing power spectral analyses on outage and restoration time series.

\subsection{Power spectral analysis \label{sec:4B}}

\begin{figure*}[!htbp]
    \centering
    \includegraphics[width=0.85\textwidth]{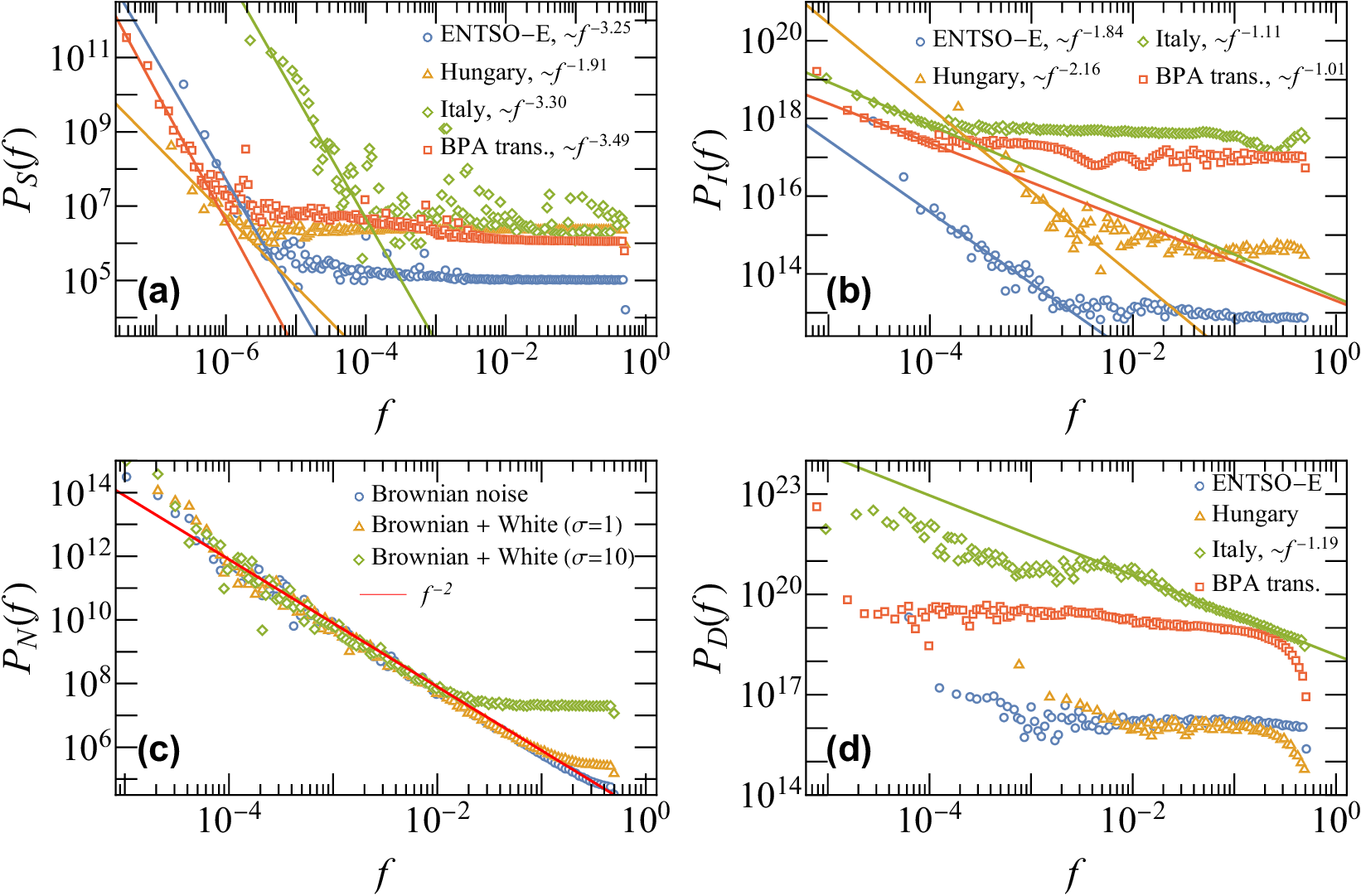}
    \caption{Power spectra of the time series of (a) the number of outage events $S(t)$, (b) the intervals between successive outage events $I(t)$, (c) the Brownian noise $N(t)$, with and without the presence of a white noise background, and (d) the outage duration $D(t)$. In panels (a), (b), and (d), we show the respective power spectra for the ENTSO-E generation outage data from all control areas, the Hungarian MAVIR generation outage data, the GME Italian generation outage data, and the BPA transmission outage data.}
    \label{fig:powsp}
\end{figure*}

In this section, we employ the spectral analysis method on the time series $S(t)$, $I(t)$, $D(t)$, and additionally $N(t)$, all introduced in Sec.~\ref{sec:2C}. As shown in Figs.~\ref{fig:powsp} (a) and (b), our collected outage data typically display a $1/f^\alpha$ power law, a signature of correlation in the outage events, in the lower-frequency range, while the higher-frequency range is overwhelmed by white noise, which is characterized by a constant power spectrum. This is immediately evident if we compare Figs.~\ref{fig:powsp} (a) and (b) to how the $1/f^2$ power spectrum of a Brownian noise is affected by white noise of different intensities, as illustrated in Fig.~\ref{fig:powsp} (c). Hence, even though the exponent values for $P_S(f)$, ranging from $\sim 1.91$ to $\sim 3.49$, are not entirely in accord with $\alpha<2$  for the Bak–Tang–Wiesenfeld and Manna sandpile models with dimensionality $d<d_c=4$ \cite{laurson2005}, we do not exclude that the topology of the system may come to play, or that there could exist a crossover in the $1/f^\alpha$ law, as the correlations manifested in the higher-frequency range are largely masked by white noise. 

Since the majority of the outage events seem to occur out of quite random causes, the above observations suggest that the manifested power laws in outage duration distributions cannot be attributed to the responses to cascade blackout events alone. In Fig.~\ref{fig:powsp} (d), we show that the outage duration time series $D(t)$ barely shows any evident traits of $1/f$ noises, except for the Italy data, in which the events were recorded hourly instead of in every minute. Since many outage events can be fixed in a matter of a few hours, we suspect that the $1/f$ noise in the high-frequency part (short time) of the Italy data merely reflects a superficial correlation of outage duration due to its low temporal resolution. Based on these arguments, the pervasive heavy tails in outage duration distributions may then be more deeply rooted in the modified HOT mechanism conjectured in \eqref{eqs:hotT}, featuring the optimized responses of the limited maintenance resources to both random outage events as well as cascade outage events resulting from SOC processes.

For outage events, the observed $1/f$ noises strengthen the complex IP + DP-C SOC hypothesis advanced before, calling for a dominating branching process for long times and a branching process overwhelmed by random events for short times and excluding the HOT explanation for outage sizes. Yet, the vastness of those random, singular events may partially be a result of the aforementioned barriers introduced to strengthen the power grids, which prevent failures from spreading further. For the $\alpha < 2$ inter-event cases, non-ergodicity, aging, and consequently an age-dependent scaling can also occur~\cite{KALASHYAN2009895}, related to so-called `crucial events'.


\section{Conclusions}
In the present paper, we revisited the topic of the size distribution of forced outages in power systems to formulate possible theoretical explanations of the uniformness of these distributions. To address a shortcoming of previous studies, long-term outage data of various power systems were collected and analyzed. First, exponents of power-law fits were extracted to cross-check the results with related literature, then this step was repeated after setting a threshold, speculating that the understanding of the manifested power law for the unavailable duration constitutes a crucial ingredient for the understanding of general power-law behavior in outage distributions. Based on the numerical results, potential explanations were presented, and a power spectral analysis was performed to demonstrate that the studied outage data are comprised of many random events as well as some correlated events characterized by the $1/f$ noise. This hints that SOC processes could take place in outage events. Therefore, for outage events, we consider the system under study as the combination of spontaneous isotropic percolation and a branching process, implying DP-C criticality. Although a SOC explanation based on the DP-C criticality for the heavy tails in outage duration distributions is tempting, given that the majority of the outage events occurred out of random causes, the manifested power laws in outage duration cannot be attributed to the responses to SOC cascading failures alone. The power spectra of the outage duration time series further indicate a lack of $1/f$ noise, leading us to conjecture an extended HOT explanation for the heavy tails in outage duration distributions. This can be quite sensible, as on the one hand, power-grid infrastructures are built more or less in a self-organized manner to meet customers' demands, so they are more rigid to give rise to SOC processes, despite some measures being introduced to confine the spread of outages; on the other hand, the needed maintenance resources in responses to outage events can be more fluidly distributed and allocated, permitting a greater extent of optimization for economical efficiency.

Since cascading failures pose higher risks to society, the statistical physics approach to 
critical phenomena opens up
the possibility to provide suggestions on how to make cascade-failure-like outages smaller.
In general, the addition of long-range interactions/correlations, 
suppresses fluctuations and increases the effective graph dimensions,
making the processes more mean-field-like, which are faster, with
shorter outage durations corresponding to larger $\tau_T$~\cite{odorbook}.
On the other hand, quenched heterogeneity and modular structures 
slow down the critical dynamics. This can be seen in our analysis
where we found smaller outage exponents for transmissions between
control areas than within them. Thus increasing the size
or decreasing the number of control areas should decrease the
outage times.
It is also known in statistical physics of critical systems 
that in multi-component models, spatial anisotropy can alter 
the universal scaling behavior~\cite{odorbook}.
In particular, for the DP-C class sand-pile models spatial 
anisotropy or an absorbing wall increases the $\tau_T$ critical 
exponent~\cite{PhysRevE.78.041102}, which could be exploited by 
deploying efficient service team geometry or topology.

In the future, it would be interesting and valuable to examine in-depth more structured data that allows a proper separation of cascading events and random outages so that one can compare how these two types of events are restored differently. If such information cannot be directly inferred from the raw data, one approach is to group outages into cascades and generations within each cascade according to their relative occurrence timing \cite{dobson2012}. In a more recent work \cite{dobsonekisheva2022}, transmission system outage data were also structured into resilience events (onset and full restoration of cascading outages) according to the timing of outages and restorations, with which several metrics were proposed to evaluate the duration of these resilience events. As power-grid systems consist of many coupled subsystems and are constantly subjected to various kinds of drive and dissipation, understanding the pervasive heavy tails in various measures is always challenging. To gain a better understanding, it will also be interesting to set up simple hybrid models with the SOC mechanism, similar to the OPA model \cite{carreras2019validating}, to account for the outage events and the HOT mechanism for optimizing the restoration processes.

\begin{acknowledgments}
Support from the Hungarian National Research, Development and Innovation 
Office NKFIH (K128989) and from the ELKH grant SA-44/2021 is acknowledged.
\end{acknowledgments}

\appendix
\section{Quantities extracted from the raw datasets \label{appenda}}
The quantities of interest are the outage duration $T_u$ and the unavailable power $P_u$. They are extracted from the raw datasets \cite{datarepo} as follows.

(i) Generator outage data:
\begin{itemize}
    \item The ENTSO-E data for different control areas in Europe \cite{entsoeprodgen}. The relevant columns of each entry are those named `avail\_qty' (available power, in unit of $\mathrm{MW}$), `end' (outage ending time), `nominal\_power' (installed power, in unit of $\mathrm{MW}$), and `start' (outage starting time). Hence, $P_u=\text{nominal\_power}-\text{avail\_qty}$ and $T_u=\text{end}-\text{start}$ (in unit of hour).

    Here we also give the definitions for a few terminologies. According to \href{https://eepublicdownloads.entsoe.eu/clean-documents/pre2015/resources/Transparency/MoP_Ref_02_-_Detailed_Data_Descriptions_v1r2.pdf}{the ENTSO-E data descriptions}, a generation unit is ``a single electricity generator belonging to a production unit'', while a production unit is understood as ``a facility for generation of electricity made up of a single generation unit or of an aggregation of generation units''. According to the document ``Commission Regulation (EU) No 543/2013 of 14 June 2013 on submission and publication of data in electricity markets'', a control area ``means a coherent part of the interconnected system, operated by a single system operator and shall include connected physical load and/or generation units if any''.
    
    \item The CAISO data for California \cite{caliiso}. We use prior trade date reports with fields for both the unavailable duration and unavailable power included. The relevant columns are those named `CURTAILMENT START DATE TIME', `CURTAILMENT END DATE TIME', and `CURTAILMENT MW': $T_u=\text{`CURTAILMENT END DATE TIME'}-\text{`CURTAILMENT START DATE TIME'}$ (in unit of $\text{hour}$), $P_u=\text{`CURTAILMENT MW'}$. 
    \item The MAVIR data for Hungary \cite{huMAVIR}. The relevant columns are those named `Start of the outage', `End of the outage', and `Unavailable capacity' (in unit of $\mathrm{MW}$): $T_u=\text{`End of the outage'}-\text{`Start of the outage'}$ (in unit of hour), $P_u=\text{`Unavailable capacity'}$.
    \item The GME data for Italy \cite{gmeitaly}. The relevant columns are those named `EventStart', `EventStop', and `UnavailableCapacity': $T_u=\text{EventStop}-\text{EventStart}$ (in unit of hour) and $P_u=\text{UnavailableCapacity}$ (in unit of $\mathrm{MW}$).
\end{itemize}

(ii) Transmission infrastructure outages:
\begin{itemize}
    \item The BPA data for the Pacific Northwest of the US \cite{BPAout}. This dataset monitored Customer Service Interruptions, Transmission Line Interruptions, and Transformer Interruptions. For the purpose of this paper, only the data for Transmission Line Interruptions and Transformer Interruptions were selected, while the data for Customer Service Interruptions was not considered because it contains blackouts for customers in different districts, which are usually consequences of multiple outage events and last as long as any infrastructure failures in a district are not restored. For the selected datasets, the relevant column is $T_u=\text{`Duration'}$ in unit of minute.
    \item The AESO historical data for Canada \cite{aesohist}. The relevant columns are `Outage Start' and `Planned End', giving $T_u=\text{`Planned End'}-\text{`Outage Start'}$ in unit of hour.
    \item The ENTSO-E data for transmission outages between pairs of control areas in Europe \cite{entsotransm}. There are 65 control areas in total \cite{datarepo}, out of which, only 226 ordered pairs of control areas have direct power exchanges. Similar to the ENTSO-E generator outage data, the relevant columns of each entry are those named `avail\_qty' (available power, in unit of $\mathrm{MW}$), `end' (outage ending time), `nominal\_power' (installed power, in unit of $\mathrm{MW}$), and `start' (outage starting time). Hence, $P_u=\text{nominal\_power}-\text{avail\_qty}$ and $T_u=\text{end}-\text{start}$ (in unit of hour). In the main text, we only show the analysis for $T_u$ in parallel with the analyses for the BPA and the AESO data. 
\end{itemize}

%

\end{document}